# An Analysis of the Relationship Between the Characteristics of Innovative Consumers and the Degree of Serious Leisure in User Innovation

Taichi Abe[1] and Yasunobu Morita[2]

[1] Graduate School of Commerce, Fukuoka University
[2] Faculty of Commerce, Fukuoka University

This study examines the relationship between the concept of serious leisure and user innovation. We adopted the characteristics of innovative consumers identified by Lüthje (2004)—product use experience, information exchange, and new product adoption speed—to analyze their correlation with serious leisure engagement. The analysis utilized consumer behavior survey data from the "Marketing Analysis Contest 2023" sponsored by Nomura Research Institute, examining the relationship between innovative consumer characteristics and the degree of serious leisure (Serious Leisure Inventory and Measure: SLIM). Since the contest data did not directly measure innovative consumer characteristics or serious leisure engagement, we established alternative variables for quantitative analysis. The results showed that the SLIM alternative variable had positive correlations with diverse product experiences and early adoption of new products. However, no significant correlation was observed with information exchange among consumers. These findings suggest that serious leisure practice may serve as a potential antecedent to user innovation. The leisure career perspective of the serious leisure concept may capture the motivations of user innovators, as suggested by Okada and Nishikawa (2019).

## 1. Introduction

In recent years, with the advancement of the information society and diversification of needs, consumer innovation has become increasingly important in the improvement and development of products and services.

Traditionally, technological progress and innovation were considered to be conducted by companies. However, among advanced users, there are lead users who experience new needs earlier than many other users while using products and services, and obtain significant benefits by solving these needs. They not only cooperate in companies' new product development but also sometimes create consumer innovations by independently developing and improving the uses and functions of products and services to meet their own needs.

This research focuses on the hobby domain, which has been significantly involved in consumer innovation both academically and practically, and particularly examines the concept of "serious leisure" commonly used in leisure research abroad. Serious leisure is the systematic pursuit of activities by amateurs, hobbyists, and volunteers, often aiming for long-term acquisition of skills, knowledge, and experience that could be called a "leisure career," providing a different perspective from traditional views of hobbies.

Through analysis using large-scale consumer behavior survey data, this study explores the possibility that serious leisure practice among sports enthusiasts is related to the characteristics of innovative consumers reported by Lüthje (2004). By doing so, we aim to provide new insights regarding the relationship with the serious leisure concept, which has not been previously examined in consumer innovation research.

# 2. Consumer Innovation in the Hobby Domain

The hobby domain is one of the areas where consumer innovation frequently occurs. Von Hippel (2019) organized a survey conducted across six countries—the UK, Japan, the US, Finland, Canada, and Korea—to identify where user innovators exist across various product categories (Table 1).

According to this research, "hobby-related" activities—combining handicrafts, gardening supplies, pet-related items, and sports/hobby items—represented the category with the highest concentration of user innovators in five countries, excluding Japan. A review of past research reveals that case studies focusing on hobby domains, particularly sports, have led to the refinement of user innovation theory and the expansion of new perspectives.

**Table 1. Percentage of User Innovation by Category in Various Countries (%)**

| Category | UK | Japan | US | Finland | Canada | Korea |
|---|---|---|---|---|---|---|
| Hobby-related | 57.0 | 24.0 | 38.6 | 37.0 | 40.0 | 34.3 |
| Housing-related | 16.0 | 45.8 | 25.4 | 20.0 | 19.0 | 17.9 |
| Children-related | 10.0 | 6.0 | 6.1 | 4.0 | 10.0 | 10.9 |
| Vehicle-related | 8.0 | 9.6 | 7.0 | 11.0 | 10.0 | 6.5 |
| Medical | 2.0 | 2.4 | 7.9 | 7.0 | 8.0 | 5.5 |
| Computer-related | na | na | na | 6.0 | 11.0 | na |
| Food/Clothing | na | na | na | 12.0 | na | na |
| Other | 7.0 | 12.0 | 14.9 | 3.0 | 3.0 | 23.9 |

Source: Created by authors based on von Hippel (2019)

Shah (2000)'s case analysis of product development in relatively new sports such as skateboarding, snowboarding, and windsurfing demonstrated the possibility that general consumers could become innovators in consumer goods categories. This was significant as previous user innovation research had primarily focused on B2B cases where companies, as users, improved products for their own use. Franke and Shah (2003) used four "extreme" sports communities (sailplane, canyoning, snowboarding cross, and handicapped cycling) as case studies to reveal that user-developed innovations were sufficiently useful and to clarify how user innovators behave within their communities. Additionally, Hienerth (2004)'s analysis of the commercialization process of user innovations in open communities included cases where kayak users created their own boats and established their own manufacturing companies when existing manufacturers wouldn't produce them.

# 3. The Concept of Serious Leisure

## 3.1. What is Serious Leisure?

Serious leisure is a concept proposed by Robert Stebbins, a Canadian leisure studies scholar. It is defined as "systematic pursuit of an amateur, hobbyist, or volunteer core activity sufficiently substantial, interesting. and fulfilling in nature for the participant to find a career there acquiring and expressing a combination of its special skills, knowledge, and experience" (Sugiyama, 2019; Stebbins, 2015). It refers to activities that are not merely pursued for relaxation but require overcoming challenges and aim to increase expertise over the long term.

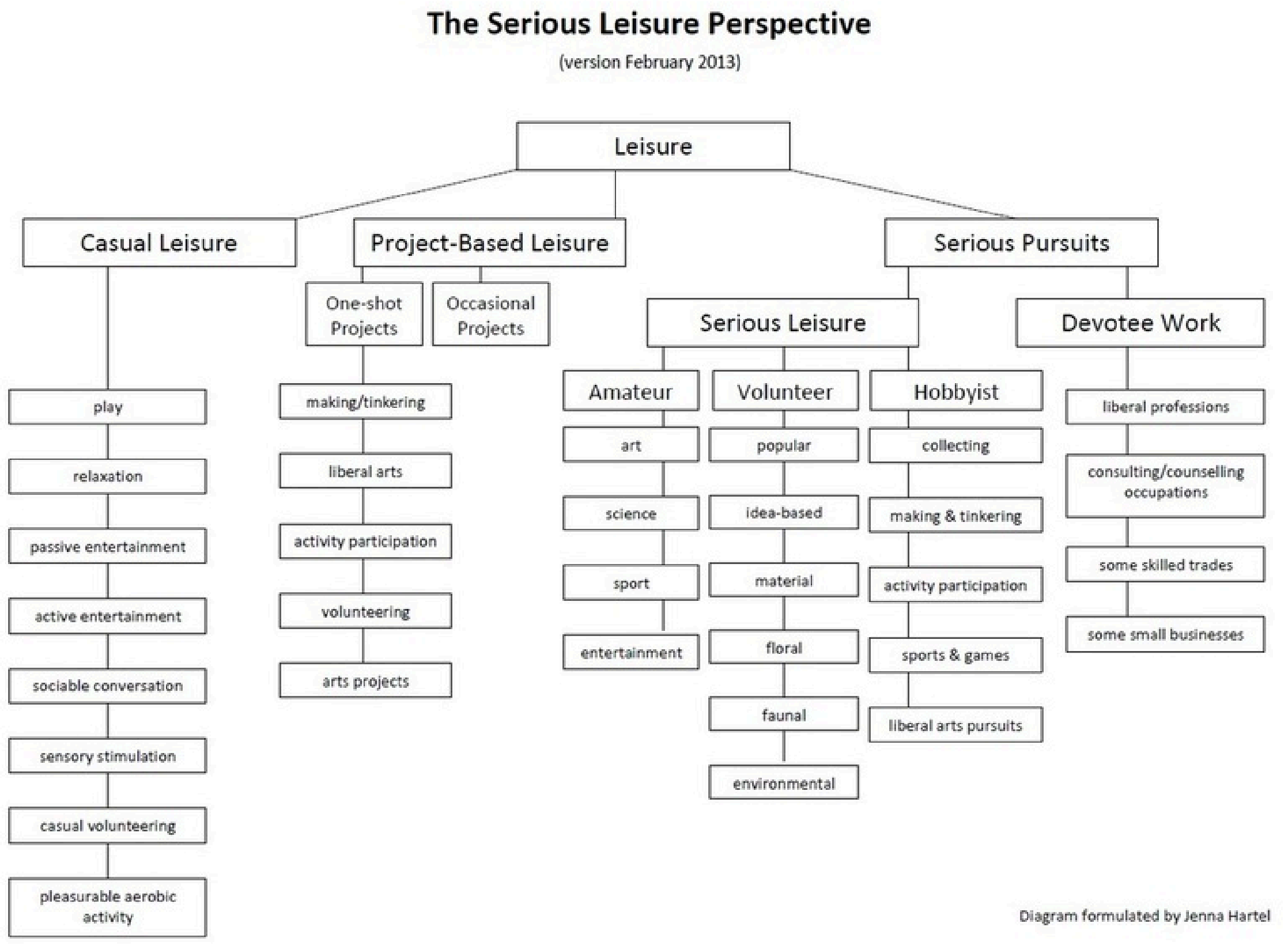

**Figure 1: SLP diagram**
**(The Serious Leisure Perspective Website, www.seriousleisure.net )**

Stebbins has systematized serious leisure as part of The Serious Leisure Perspective (SLP) framework, along with "casual leisure," which can be enjoyed immediately without special training, and "project-based leisure," which involves moderately complex creative attempts that occur occasionally but don't form a leisure career. He theoretically argues that all leisure activities can be organized into these three forms (Figure 1).

The serious leisure concept has been widely studied, particularly in Western sociology. Sugiyama (2019) conducted a narrative review of 70 serious leisure-related articles published in three major Western academic journals in leisure research through 2018. This analysis revealed that past serious leisure research could be classified into eight major categories: "clarification of social worlds," "clarification of constraint coping strategies," "clarification of gendering," "clarification of minority practices," "clarification of effects on quality of life," "illumination of serious leisure aspects," "theoretical refinement," and "examination of relationships with adjacent concepts." This classification demonstrates that serious leisure research has been pursued from multiple angles.

## 3.2. Serious Leisure Measurement Scales

Regarding the quantitative measurement of serious leisure, several scales have been developed, but the most widely recognized is the Serious Leisure Inventory and Measure (SLIM) developed by Gould et al. (2008). This scale sets 18 dimensions based on the six characteristics of serious leisure that Stebbins frequently references: "persevere", "leisure career", "effort", "durable benefits", "unique ethos", and "identity" (Stebbins, 1992; 2015). It measures to what extent an individual practices an activity as serious leisure.

Gould et al. (2008) proposed both the original 72-item version and a shortened 54-item version created by removing the item with the lowest factor loading from each dimension's four items. Furthermore, Gould et al. (2011) proposed an 18-item short-form scale as a useful and less burdensome measure by extracting the items with the highest factor loadings from each dimension of the 54-item SLIM in a survey of chess players.

Lee et al. (2023) conducted a comprehensive review of SLIM's applications and usefulness. They analyzed 34 academic papers published in English and revealed that SLIM was primarily used for two purposes: ① classification of serious leisure practitioners and ② verification of relationships between serious leisure and other psychological and behavioral variables. These findings support SLIM as a useful and versatile measurement tool in serious leisure research.

## 3.3. Potential Conceptual Relationship Between Serious Leisure and Consumer Innovation

Von Hippel (2006) points out that individual user innovators gain benefits not only from the developed products themselves but also "from the process of developing or modifying products" (p.86). Jeppesen & Frederiksen (2006) revealed that innovative users who contribute to corporate user communities are hobbyists who are not motivated by monetary rewards. Furthermore, Honjo (2022) demonstrated that the fields and means through which individual consumers realize innovations are closer to hobby domains than academic or professional fields, stating that "consumer innovation is primarily realized in fields close to hobbies using means that are close to hobbies" (p.105).

Moreover, as serious leisure characteristics include "perseverance" and "significant effort," serious leisure is not merely a pastime or entertainment but a way of engaging in hobbies that pursues unique value important to individuals and requires effort to achieve challenging goals. Therefore, people who engage in a hobby as serious leisure are likely to have high requirements for products and services related to that hobby and may actively engage in improving existing products or developing new ones.

Thus, while hobbies practiced by individuals play a crucial role in realizing consumer innovation, and there appear to be many conceptual connections between consumer innovation and serious leisure, few studies link these two concepts. This research aims to provide a new perspective on the relationship between serious leisure concepts and consumer innovation research through the analysis of large-scale consumer behavior data.

# 4. Research Method

## 4.1. Data Used

This research utilizes data provided by the "Marketing Analysis Contest 2023" sponsored by Nomura Research Institute. The survey was conducted between January and March 2023 on 2,500 people primarily in the Kanto region, including items asking about participation in 32 hobbies. This study analyzed samples (n = 609) who answered "yes" to at least one of five hobby-related questions: "sports/fitness," "golf," "skiing/snowboarding," "fishing," and "outdoor activities/camping."

The purpose of this study is to clarify the relationship between serious leisure and consumer innovation. Since the data did not include items that directly measure serious leisure, we used the sum of promotion focus items 2 and 5 (7-point scale) from the regulatory focus scale (Table 3). These two question items were similar in content to the "leisure career" and "significant effort" measurement items in SLIM, and were therefore treated as alternative measurements for SLIM. Hereafter in this study, we refered to the sum of regulatory focus scale promotion focus items 2 and 5 as the "serious leisure score"[1].

## 4.2. Alternative Variables and Hypothesis Development

Lüthje (2004) identified four characteristics of innovative consumers who create consumer innovations in outdoor sports products: ① product use experience, ② information exchange with other users, ③ technical expertise, and ④ speed of new product adoption.

This study examined how these characteristics (Table 4) relate to the serious leisure score of people who engage in sports as a hobby.

Based on this, we set the following hypotheses to investigate the relationship between serious leisure scores and innovative consumer characteristics. These hypotheses, H1 through H3, were set regarding the characteristics of consumers with high serious leisure scores among those who engage in sports and outdoor activities as hobbies.

---

[1] From the perspective that appropriate question items should be used for the construct being measured, this method may not be entirely appropriate when compared to the regulatory focus theory, which has accumulated much research since Higgins (1997). However, what this study aims to explore is whether the serious leisure concept can be applied to the fields of consumer behavior and marketing. Therefore, we adopted this as an alternative method utilizing regulatory focus theory (e.g., Iino 2018), which is becoming established in marketing and consumer behavior fields.

H1: Consumers with higher serious leisure scores have more various sports experiences.

H2: Consumers with higher serious leisure scores engage in more information exchange with other consumers.

H3: Consumers with higher serious leisure scores have faster new product purchase and adoption speeds.

## 4.3. Analysis Overview

The analysis used R (4.2.0). After calculating descriptive statistics and Cronbach's alpha coefficient, we calculated Pearson's correlation coefficients between serious leisure score (variable name: SL_score) and "various sports experiences" (sum_sports), "information exchange with other consumers" (sum_comm), and "new product purchase/adoption speed" (REC_Scale_11_MX). This evaluated the strength and direction of relationships between each variable and serious leisure scores. To confirm the statistical significance of correlation coefficients, we conducted significance tests (t-tests) for each correlation coefficient. The significance level was set at 5% ($\alpha = 0.05$), and two-tailed tests were performed.

**Table 2. Descriptive Statistics**

| Variable | Minimum | Maximum | Mean | SD |
|---|---|---|---|---|
| Serious Leisure Score | 2 | 14 | 9.0 | 2.4 |
| Various Sports Experiences | 1 | 5 | 1.4 | 0.7 |
| Information Exchange with Other Consumers | 0 | 2 | 0.3 | 0.5 |
| New Product Purchase/Adoption Speed | 1 | 5 | 2.5 | 1.2 |

**Table 3. Comparison of Serious Leisure Measurement Scale (Gould et al., 2008) and Regulatory Focus Scale Items (Hows et al., 2010; Ishii, 2020)**

| SLIM:<br>Serious Leisure Inventory and Measure | Regulatory Focus Scale<br>- Promotion Focus Items |
|---|---|
| Career Progress<br>4. I have progressed in __ since beginning. | 2.これまでに、自分の人生における成功に向かって前進してきたと感じる。<br>(I feel like I have made progress toward being successful in my life.) |
| Effort<br>4. I am willing to exert considerable effort to be more proficient at __. | 5. 自分は、希望・願い・憧れを実現して「理想的な自分」に近づけるよう積極的に努力する人間であると思う。<br>(I see myself as someone who is primarily striving to reach my “ideal self”—to fulfill my hopes, wishes, and aspirations.) |

**Table 4. Correspondence between Innovative Consumer Characteristics (Lüthje, 2004) and Variables Used**

| Innovative Consumer Characteristics | Alternative Variables |
|---|---|
| ① Use Experience (Number of different sports) | Total score of "yes" (=1) responses for hobbies "sports/fitness," "golf," "skiing/snowboarding," "fishing," and "outdoor activities/camping" used as "various sports experiences" |
| ② Information exchange with other sportsmen | Total score (Cronbach's alpha = 0.18) of "yes" (=1) responses to consumption value items "concerned about users' opinions" and "actively share information about products and stores" |
| ③ Technical know-how | No suitable alternative items available |
| ④ Speed of adoption | REC scale (5-point Likert scale) "buy new things earlier than others" |

# 5. Results

## 5.1. Descriptive Statistics and Scale Properties

Descriptive statistics are shown in Table 2. Floor effects were observed for "various sports experiences" and "information exchange with other consumers." For the former, most consumers tended to have experience in approximately one type of sport, and there was a bias toward little information exchange.

The Cronbach's alpha coefficient for "serious leisure score," calculated from the sum of two variables, was 0.73, showing acceptable internal consistency.

## 5.2. Correlation Analysis Results

The results of the correlation analysis are shown in Table 5.

**Table 5. Correlation Coefficients and Significance between Serious Leisure Score and Each Variable**

| Variable | r | t-value | p-value |
|---|---|---|---|
| Various sports experiences (sum_sports) | 0.21 | 5.23 | <0.01 |
| Information exchange with other consumers (sum_comm) | 0.02 | 0.55 | 0.58 |
| New product purchase/adoption speed (REC_Scale_11_MX) | 0.35 | 9.08 | <0.01 |

First, a weak positive correlation ($r = 0.21$, $p < .01$) was found between various sports experiences (sum_sports) and serious leisure score. This result was statistically significant, and hypothesis H1 was supported.

Regarding information exchange with other consumers (sum_comm), no statistically significant correlation was found with serious leisure score ($r = 0.02$, $p = 0.58$). Therefore, hypothesis H2 was rejected.

A moderate positive correlation ($r = 0.35$, $p < .01$) was found between new product purchase/adoption speed (REC_Scale_11_MX) and serious leisure score. This result was statistically significant, and hypothesis H3 was supported.

# 6. Discussion

Of the three hypotheses, H1 and H3 were supported. Although the correlation coefficients were small, the results suggest a possible relationship between innovative consumer characteristics and serious leisure scores among people who engage in sports and outdoor activities as hobbies. However, this study has several limitations in hypothesis setting and operational definition of variables.

## 6.1. H1: "Having Various Sports Experiences"

While hypothesis H1 was supported, its correlation coefficient was very weak. In the "various sports experiences" variable used for verification, the "sports/fitness" option in the survey data encompasses various competitions and activities. Meanwhile, "golf," "skiing/snowboarding," "fishing," and "outdoor activities/camping" are asked about individually. Due to this structure, the characteristics of samples engaging in these four activities/competitions may have a significant influence on the overall variable. Such bias might affect the validity of the "various sports experiences" variable and may not accurately reflect innovative consumer characteristics. Consequently, the relationship between this variable and serious leisure scores requires careful interpretation.

## 6.2. H2: "Engaging in More Information Exchange with Other Consumers"

We could not confirm a positive correlation between information exchange with other consumers and serious leisure scores for hypothesis H2. Additionally, the Cronbach's alpha coefficient value for the "information exchange with other consumers" variable used to test hypothesis H2 was 0.18, not showing internal consistency.

Honjo (2016) introduces two concepts representing consumer interaction regarding new product diffusion: "opinion leadership," which represents the degree of influencing new product adoption by providing information to other users, and "opinion seeking," which represents the degree of seeking information from other users when adopting new products themselves (Flynn, Goldsmith, and Eastman, 1996). Regarding these two concepts, Schreier et al. (2007) revealed that the degree of lead users (using leading edge status here) positively influences opinion leadership but negatively influences opinion seeking.

On the other hand, the "information exchange with other consumers" variable used to test hypothesis H2 was created by combining two items: the opinion seeking-like SEN_10_MA "concerned about users' opinions" and the opinion leader-like SEN_26_MA "actively share information about products and stores." In a preliminary analysis examining correlations between SEN_10_MA and SEN_26_MA individually with serious leisure scores, the opinion seeking-like value SEN_10_MA showed no significant correlation with serious leisure scores ($r = -0.06$, $p = 0.17$). The opinion leader-like value SEN_26_MA showed a very small but positive correlation with serious leisure scores (0.12) and met statistically significant levels ($p < .001$). The variable used to test hypothesis H2 in this study may not have captured innovative consumer characteristics as it combined potentially opposing effects of opinion seeking and opinion leadership.

Furthermore, Franke and Shah (2003) reported that sports communities with intense competition between members showed considerably less willingness to share innovation-related information compared to other sports communities. In this survey, the activities/competitions practiced by the sample respondents might include many that involve competing in individual skills or performance, potentially affecting the relationship with serious leisure scores. However, as dimensions like "Group Attraction" and "Group Maintenance" are included in SLIM, the practice of serious leisure cannot be separated from the formation and maintenance of enthusiast communities. People who participate in serious leisure communities are likely to frequently exchange information about related products through their social connections. How information about related products is exchanged within serious leisure communities and how this relates to consumer innovation remains a topic that should be clarified through continued literature review and survey implementation.

### 6.3. H3: "Faster New Product Purchase and Adoption Speed"

Hypothesis H3 was supported by a statistically significant moderate positive correlation coefficient. This result suggests a relationship between serious leisure scores and early adoption tendencies for new products. According to Honjo's (2016) review, previous research has shown that early adoption behavior of new products can be consistently explained by lead userness. The findings of this study suggest that serious leisure practitioners may have characteristics similar to lead users, raising the need to further explore the relationship between these two concepts.

## 7. Conclusion

The results of this study suggest that serious leisure practitioners may act as consumer innovators for products related to their hobbies. Future research should examine whether serious leisure and consumer innovation are related in hobbies beyond sports. Moreover, while it is known that lead user factors include both domain-specific factors and domain-independent factors such as internal control, innovativeness, and divergent thinking (Honjo, 2016), detailed examination through both quantitative and qualitative approaches is needed to determine whether the practice of hobbies as serious leisure relates to domain-independent factors, although its relationship with domain-specific factors like product knowledge and usage experience is readily conceivable.

Traditionally, the diffusion of tangible products, unlike intangible products such as software, was considered impossible to complete within user communities alone (von Hippel, 2006,

p.164-165). However, in recent years, Fab spaces equipped with fabrication tools such as 3D printers and laser cutters have been established worldwide, popularizing means for general consumers to develop, improve, and distribute tangible products. The existence of distinctive values called the maker movement, which is expanding through this trend, suggests that people participating in these communities practice their hobbies as serious leisure. The perspective of serious leisure may capture the transition in motivations of consumers with making hobbies who drive the maker movement, as pointed out by Okada & Nishikawa (2019).

This research also has practical implications. While the usefulness of the lead user method in product development is widely recognized, efficiently identifying lead users for one's products remains a challenge. Conducting surveys from the new perspective of serious leisure might lead to developing approaches for discovering lead users.

Furthermore, serious leisure research addresses the exploration of people's motivations and values in hobby activities. Therefore, serious leisure research is expected to provide valuable implications not only for innovation research but also for broader fields of commercial research, including marketing and consumer behavior studies.

Note: This research is a revised and expanded version of an entry submitted to the "Marketing Analysis Contest 2023" sponsored by Nomura Research Institute.